\documentclass[12pt]{article}
\usepackage{amsmath}
\usepackage{graphicx,psfrag,epsf}
\usepackage{enumerate}
\usepackage{natbib}
\usepackage{url} % not crucial - just used below for the URL 
\usepackage{bm}
\usepackage[utf8]{inputenc}
\usepackage[english]{babel}
\usepackage{amsthm}
\usepackage{url}
\usepackage{amsfonts}
\newtheorem{theorem}{Theorem}
\usepackage{multirow}
\usepackage{xcolor}
\newcommand{\code}[1]{\texttt{#1}}
\newcommand{\blind}{1}
\usepackage{multirow}
\usepackage{comment}
\newtheorem{remark}{Remark}

\begin{document}

\def\spacingset#1{\renewcommand{\baselinestretch}%
{#1}\small\normalsize} \spacingset{1}

%%%%%%%%%%%%%%%%%%%%%%%%%%%%%%%%%%%%%%%%%%%%%%%%%%%%%%%%%%%%%%%%%%%%%%%%%%%%%%

\if1\blind
{
  \title{\LARGE\bf A Joint Spatial Conditional Auto-Regressive Model for Estimating HIV Prevalence Rates Among Key Populations}
  \author{Zhou Lan
  %\thanks{The authors gratefully acknowledge \textit{please remember to list all relevant funding sources in the unblinded version}}\hspace{.2cm}
  \\
    School of Medicine, Yale University\\
    and \\
    Le Bao \\
    Department of Statistics, Penn State University}
  \maketitle
} \fi

\if0\blind
{
  \bigskip
  \bigskip
  \bigskip
  \begin{center}
    {\LARGE\bf A Joint Spatial Conditional Auto-Regressive Model for Estimating HIV Prevalence Rates Among Key Populations}
\end{center}
  \medskip
} \fi

\bigskip
{\LARGE\bf }

\bigskip
\begin{abstract}
Ending the HIV/AIDS pandemic is among the Sustainable Development Goals for the next decade. In order to overcome the gap between the need for care and the available resources, better understanding of HIV epidemics is needed to guide policy decisions, especially for key populations that are at higher risk for HIV infection. Accurate HIV epidemic estimates for key populations have been difficult to obtain because their HIV surveillance data is very limited. In this paper, we propose a so-called joint spatial conditional auto-regressive model for estimating HIV prevalence rates among key populations. Our model borrows information from both neighboring locations and dependent populations. As illustrated in the real data analysis, it provides more accurate estimates than independently fitting the sub-epidemic for each key population. In addition, we provide a study to reveal the conditions that our proposal gives a better prediction. The study combines both theoretical investigation and numerical study, revealing strength and limitations of our proposal.

\end{abstract}

\noindent%
{\it Keywords:}  Conditional Auto-Regressive Model, Cross-Population Dependence, HIV Prevalence, Key Populations, Missing Data
\vfill

\newpage
\spacingset{1.45} % DON'T change the spacing!
\section{Introduction}
\label{sec:int}

% , based on sexual practices, occupations, and substance use
Almost four decades since the HIV/AIDS pandemic began, HIV continues to be a leading cause of death \citep{Naghavi2017global}. Ending the HIV epidemic is among the Sustainable Development Goals (\url{https://sustainabledevelopment.un.org/}) for the next decade \citep{Alfven2017global,Bekker2018advancing,WHO2019global}. However, it is challenged by the long-standing gap between the need for care and the available resources to provide care for the populations mostly affected by the HIV epidemic. These populations are called \textit{key populations}, and they are at higher risk for HIV, based on sexual practices, occupations, and substance use (e.g., injection drug use, female sex workers, and men who have sex with men) \citep{lyerla2008quality,calleja2010has,baral2012burden}. Accurate HIV epidemic estimates among key populations would help determine the governments' policy and resource allocation. To monitor the HIV epidemics among key populations, countries rely on anonymous HIV surveillance data which include the sample size of participants and the proportion of HIV positive cases that are collected at sexually transmitted disease (STD) clinics. In most countries, the HIV surveillance data for key populations is still very limited. In light of this, a model that more efficiently utilizes existing data and produces more accurate estimates of HIV prevalence among key populations is needed\footnote{The surveillance data for the remaining population (the population which is not a \textit{key population}) are relatively abundant at clinics, and thus estimating the HIV prevalence among the remaining population is not the focus of this paper.}.

The generalized {linear} mixed model is an appealing tool for information pooling. One may assume that the number of HIV positive cases follows a binomial distribution with the unknown proportion parameter corresponding to the HIV prevalence within a key population at a certain time and location. The fixed effects and the the random effects are specified accordingly to capture the population-level effects and potential randomness across key populations and clinics. For instance, the most widely used estimates of HIV prevalence and incidence trends are created by statistically fitting the mixed effects model to HIV surveillance data \citep{Bao2012modelling,Niu2017incorporation,Eaton2019estimation}. The random effects are clinic-specific and thus were assumed to be independently distributed; and the key populations were modeled separately. However, we may conjecture that the prevalence rates of the relevant populations can be jointly high/low at the same location and the same year because of the HIV transmission pathway, and incorporating this cross-population dependence may induce more accurate estimates. In addition, the prevalence rates of the remaining people may act as offsets indicating the variation of all other key populations \citep{spiegel2004hiv,eaton2011concurrent}. {To induce both the spatial dependence and the cross-population dependence, our proposed model is constructed with the following figures: (1) the spatial conditional auto-regressive (CAR) model \citep{besag1974spatial} captures the spatial dependence; (2) the cross-population dependence assumes that the prevalence rates of any two populations at a location in the same year are correlated. The approach we construct cross-population dependence was known as \textit{Gaussian cosimulation} and was introduced by \citet{oliver2003gaussian}. It has a variety of applications, e.g.,  environment \citep{recta2012two}, ecology \citep{fanshawe2012bivariate}, portfolio analysis \citep{weatherill2015exploring}, etc. We implement this model in a Bayesian way and use the posterior predictive distribution to impute the missing entries of the key populations. We obtain a substantial improvement in imputation accuracy on real HIV prevalence datasets.}

{The availability of HIV surveillance is extremely imbalanced for key populations -- some locations have data for all key populations while some locations do not have any key population data. We provide an investigation of missing structures to understand when our proposed model would be expected to yield improved results. Essentially, the complicated missingness of surveillance data can be categorize into two types: \textit{matching} and \textit{discrepancy}. We study the impact of two missing structures on the parameter estimation and missing data imputation, and find an interesting trade-off between two missing structures.} 

In the rest of the paper, we first introduce our motivating data in Section \ref{sec:data} and our method in Section \ref{sec:Method}. In light of our scientific goal, which is to impute the missing HIV prevalence among different key populations, we use our motivating data to evaluate our proposal via measuring the accuracy in imputation (Section \ref{sec:Applications}). Given that our proposal provides more accurate imputations, we further investigate the missing structure trade-off (Section \ref{sec:Simulation}). In Section \ref{sec:Conlusion}, we conclude with a discussion.

\section{HIV Prevalence Data}
\label{sec:data}

%\textcolor{red}{This is old text:} In this paper, we use the HIV surveillance data from three representative countries to demonstrate our proposal. They are Ukraine (2005-2015), Morocco (2001-2018), and Jamaica (2000-2014).  We have the number of HIV-infected subjects and the sample size of participants for each population, i.e., injection drug use (IDUs) \textcolor{red}{(people who use drugs (PWID) is the new standard name used by WHO)}, female sex workers (FSW), sex clients (Clients), and/or men who have sex with men (MSM), and remaining people of a country's top level of division during recent years. We use integers to denote a population, a location, or a year. Here, we take Ukraine as an example: $i=1:5$ represents IDUs, FSW, Clients, MSM, and remaining people, respectively; $j=1:17$ represents the districts (i.e., Kiev, Crimea, etc), respectively; $k=1:12$ represents the year from 2004-2015, respectively. Let $Y_{ijk}$ and $N_{ijk}$ be the number of HIV infected people and its associated sample size of the population $i\in\{1,2,...,I\}$ at the location $j\in\{1,2,...,J\}$ in the year $k\in\{1,2,...,K\}$, respectively.  

\begin{figure}[ht!]
    \centering
    \includegraphics[width=0.9\textwidth]{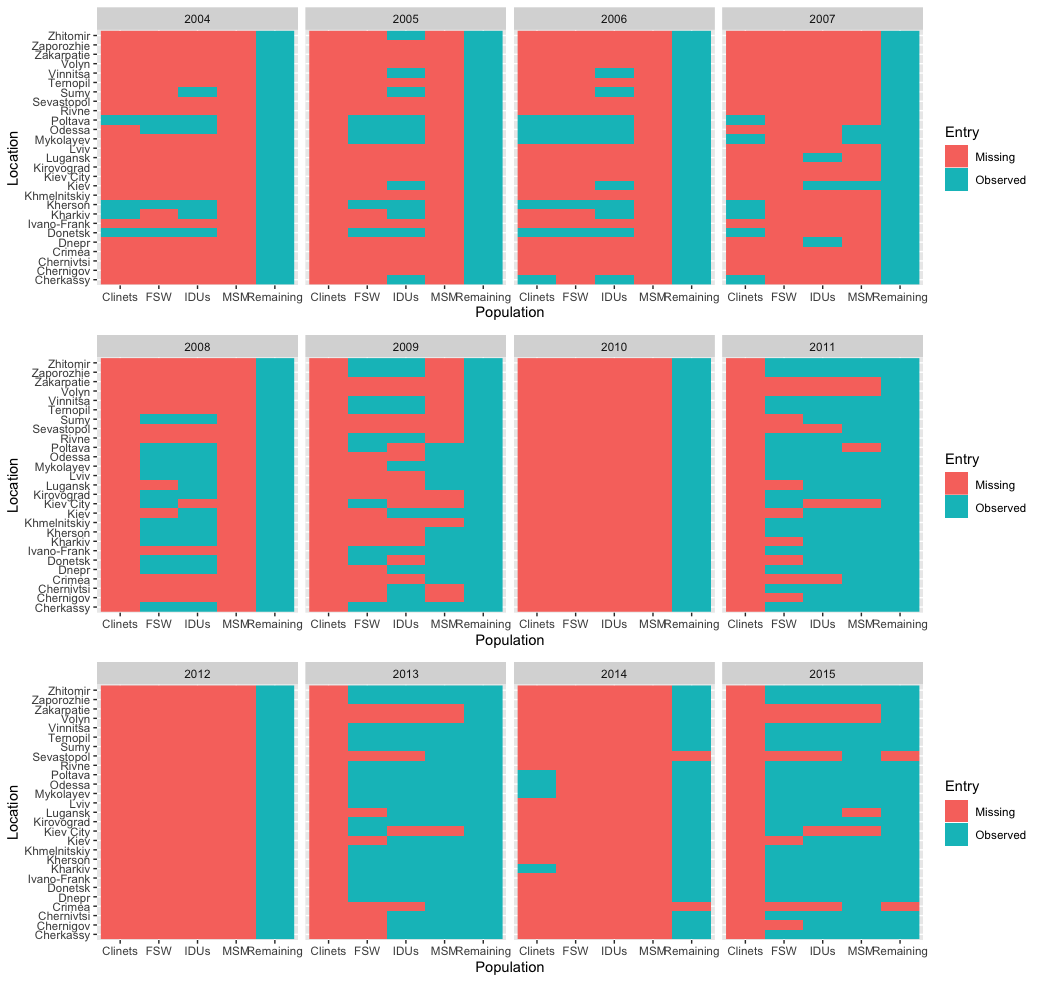}
    \caption{The missing structure of Ukraine from 2004 to 2015 are visualized. The y-axis is for the populations and the x-axis is for the locations. The red entries are missing and the blue entries are observed.}
    %\textcolor{red}{How about gray for observed and white for missing? It will save inks for reviewers, and printed red and green can not be distinguished.}
    \label{fig:UKR_missing}
\end{figure}

{In this paper, we use the HIV surveillance data from three representative countries to demonstrate our proposal. They are Ukraine (2004-2015), Morocco (2001-2018), and Jamaica (2000-2014). We use subscripts, $i$, $j$, $k$, to denote a population, a location, and a year, respectively; and we let $Y_{ijk}$ be the number of HIV positive cases and $N_{ijk}$ be the sample size of the population $i\in\{1,2,...,I\}$ at the location $j\in\{1,2,...,J\}$ in the year $k\in\{1,2,...,K\}$, respectively.} {Take Ukraine data as an example: $i=1:5$ represent people who use drugs (IDUs), female sex workers (FSW), clients of female sex workers (Clients), men who have sex with men (MSM), and remaining people; $j=1:27$ represent the districts (i.e., Kiev, Crimea, etc); $k=1:12$ represent the year from 2004-2015. We can easily provide naive prevalence rate estimators $\hat{p}_{ijk}=\frac{Y_{ijk}}{N_{ijk}}$ for the combinations of $(i,j,k)$ if the surveillance data is available. However, except for the remaining people, the observations of the key populations suffer severe data scarcity. Figure \ref{fig:UKR_missing} gives the missing structure of Ukraine. As stated before, the missingness is due to the limitation of the HIV surveillance data for key populations, and thus the corrasponding $Y_{ijk}$ and $N_{ijk}$ are not observed. Our primary goal is to provide the HIV prevalence estimates for all key populations.}

%\textcolor{red}{This is old text:} The samples at Ukraine's top level of divisions (24 oblasts, 1 autonomous republic, and 2 cities with special status) are collected every year from 2004-2015. \textcolor{red}{There is an inconsistency between 17 and 24. We can simply remove this sentence instead of further explain the details.} Figure \ref{fig:UKR_missing} gives the missingness due to the limitation of the HIV surveillance data for key populations. Except for the remaining population, the observations of the key populations suffer severe data scarcity. Obtaining the prevalence rates is our primary goal. We can easily provide naive prevalence rate estimators $\tilde{p}_{ijk}=\frac{Y_{ijk}}{N_{ijk}}$ for the combinations of $(i,j,k)$ if the surveillance data is available. However, for the entries where the surveillance data is unavailable, we need togive inferred prevalence estimates for the combinations.  

\section{Method}
\label{sec:Method}
In this section, we give our proposed model. As introduced in Section \ref{sec:data}, we let $Y_{ijk}$ and $N_{ijk}$ be the number of HIV infected people and the sample size of participants, respectively. Since $Y_{ijk}$ is the number of infected people among $N_{ijk}$ participants of the population $i$ at the location $j$ in the year $k$, we assume that $Y_{ijk}$ follows a binomial distribution with the HIV prevalence rate $p_{ijk}$, denoted as
$$Y_{ijk}|p_{ijk}\sim \mathcal{B}(N_{ijk},p_{ijk}).$$
We use the logit link function as the link function, and thus we have the transformed mean as $\mu_{ijk}=\log\frac{p_{ijk}}{1-p_{ijk}}$. Under the framework of linear mixed model, we assume that the variation of the transformed mean $\mu_{ijk}$ can be decomposed into a fixed effect $v_i(k)$ describing the population specific time trend which is a quantity of interest, and a random effect $s_{ij}$ describing the additional variability across populations and locations. This decomposition is expressed as $$\mu_{ijk}=v_i(k)+s_{ij}.$$

\subsection{Effect Specification}
\begin{figure}[ht!]
    \centering
    \includegraphics[width=1\textwidth]{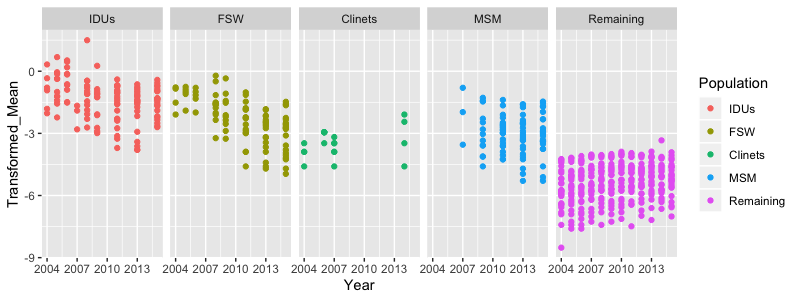}
    \caption{The {population-specific HIV prevalence} trends of Ukraine from 2004 to 2015 are visualized by using scatter plots. The y-axis is for the transformed ($\hat{\mu}_{ijk}$) and the x-axis is for the years.}
    \label{fig:Trend_UKR}
\end{figure}

First, we give the specification of the fixed effect. In our model, the fixed effect is defined as an effect driven by the population-specific trend over the years, describing the averaged level of the prevalence rate in a certain year ($k$). The trend is usually dynamic and population-specific. For example, Figure \ref{fig:Trend_UKR} indicates that the transformed means ($\hat{\mu}_{ijk}=\log\frac{\hat{p}_{ijk}}{1-\hat{p}_{ijk}}$) present different trends among populations and are non-trivial to be handled. Among a variety of nonparametric regression models and given our epidemic real data, we use the cubic-polynomial regression to model the population-specific trend, that is that $$v_i(k)=\beta_{0i}+\sum_{p=1}^3 k^p\beta_{ki}.$$

\begin{figure}[ht!]
    \centering
    \includegraphics[width=0.9\textwidth]{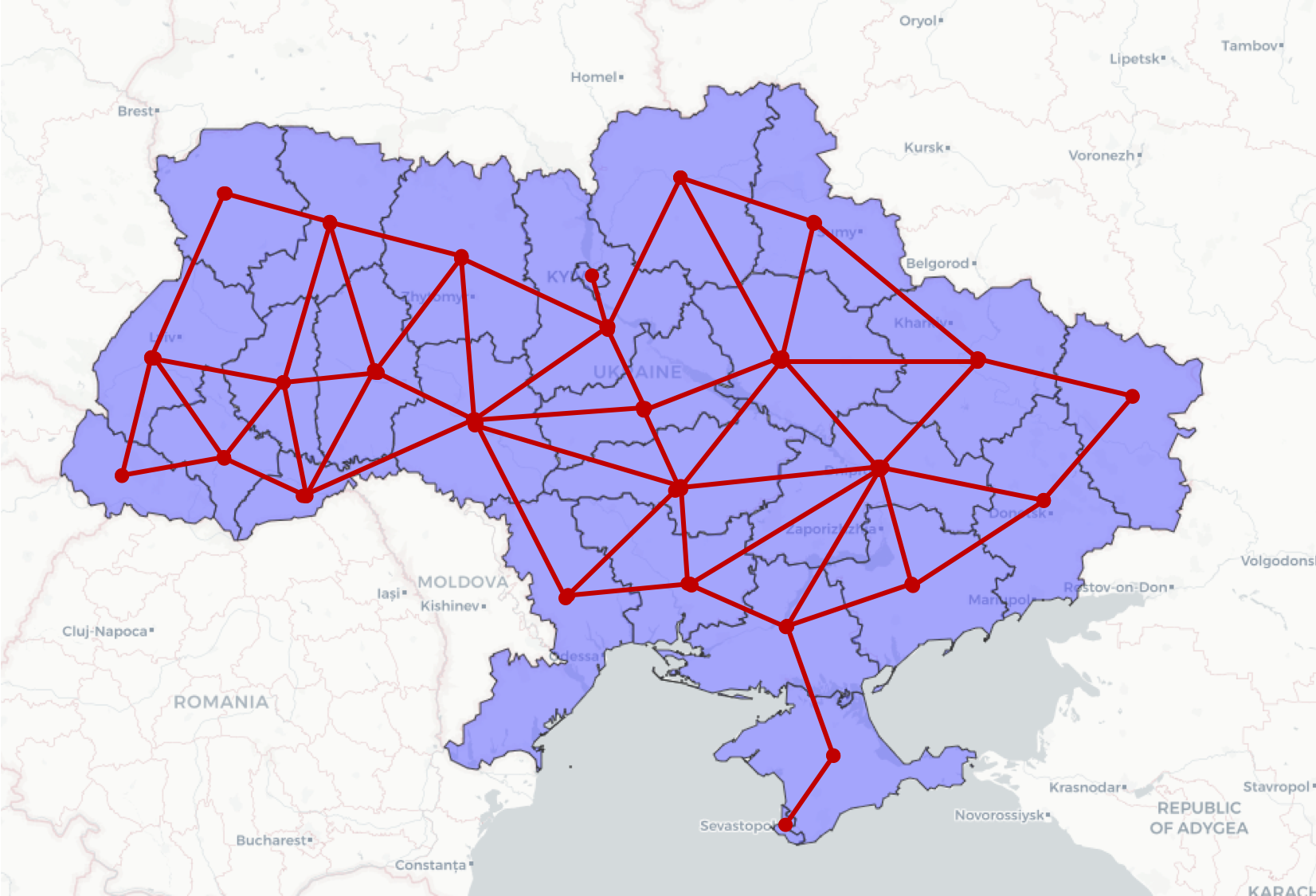}
    \caption{The administrative map of Ukraine. The nodes are connected if the locations are adjacent.}
    \label{fig:map_UKR}
\end{figure}

Next, we introduce the specification of the random effect $s_{ij}$. There are several potential options for this specification. The random effect $s_{ij}$ can be simply treated as an effect caused by the variation among locations, e.g., $s_{ij}\sim\mathcal{N}(0,\sigma^2_i)$. However, this specification does not allow the spatial dependence among nearby locations. Given that the HIV prevalence data can be treated as areal data \citep{rue2005gaussian}, another popular approach is the spatial conditional auto-regressive (CAR) model \citep{besag1974spatial}. In the CAR model, we treat the locations as the nodes of an undirected graph and the nodes are connected if the locations are adjacent (See Figure \ref{fig:map_UKR}). The random effects are specified as $$\bm{s}_i=[s_{i1}, s_{i2}, ..., s_{iJ}]^T\sim \mathcal{N}(\bm{0},\sigma_i^2\bm{D}(\bm{I}-\phi_i\bm{C})^{-1}),$$
where $\bm{D}$ is a diagonal matrix whose diagonal entries are the degrees of each node and $\bm{C}$ is the adjacency matrix of the graph\footnote{In graph theory, an undirected graph is made up of nodes (locations) which are connected by edges. In this context, degrees of a node is the number of the other nodes which are connected to it. Adjacency matrix is a $n\times n$ symmetric matrix. The entries of the matrix can only be $0$ or $1$. If node $i$ and node $j$ are connected, the $(i,j)$-th and $(j,i)$-th entry are 1; otherwise, it is $0$.}. The population-specific variance $\sigma^2_i>0$ and the population-specific spatial parameter $0<\phi_i<1$ control the local variance and the spatial dependence, respectively. 

The CAR model only accounts for spatial dependence but not cross-population dependence. The so-called cross-population dependence is defined as joint variability between any two populations, e.g., the prevalence rates of the sex workers and their clients are jointly high/low. An approach to handle this variability is to assume that the covariance matrix between $\bm{s}_{i}$ and $\bm{s}_{i'}$ is $\rho_{ii'}\bm{L}_i\bm{L}_{i'}^T$, where $\rho_{ii'}=\rho_{i'i}\in [-1,1]$, $\bm{\Sigma}_{i}=\sigma_i^2\bm{D}(\bm{I}-\phi_i\bm{C})^{-1}$ and its Cholesky decomposition is $\bm{\Sigma}_{i}=\bm{L}_i\bm{L}_i^T$. Thus, the dependence between the population $i$ and the population $i'$ is determined by $\rho_{ii'}$: the two populations have large positive dependence if $\rho_{ii'}$ is close to $1$; the two populations have large negative dependence if $\rho_{ii'}$ is close to $-1$; and the two populations have no dependence if $\rho_{ii'}$ is close to $0$. This approach is referred to as \textit{Gaussian cosimulation}, and the validity of the above approach for constructing cross-covariances has already been provided \citep{oliver2003gaussian}. \citet{recta2012two} further gives an intuitive explanation of $\rho_{ii'}$, that is, the correlation between the processes of the population $i$ and $i'$ at the same location.

\subsection{A Joint Spatial Conditional Auto-Regressive Model}
To this end, our proposed model is 
\begin{equation}
\label{eq:proposed}
    \begin{aligned}
     &Y_{ijk}|p_{ijk}\sim \mathcal{B}(N_{ijk},p_{ijk}),\   \mu_{ijk}=\log\frac{p_{ijk}}{1-p_{ijk}}=v_i(k)+s_{ij}\\
     &v_i(k)=\beta_{0i}+\sum_{p=1}^3 k^p\beta_{ki},\   [\bm{s}_{1}^T, \bm{s}_{2}^T, ..., \bm{s}_{I}^T]^T\sim \mathcal{N}(\bm{0},\bm{S})\\
     &\bm{S}=\begin{bmatrix}
    \bm{\Sigma}_1 & \rho_{12}\bm{L}_1\bm{L}_2^T & \dots & \rho_{1I}\bm{L}_1\bm{L}_I^T \\
    \rho_{21}\bm{L}_2\bm{L}_1^T & \bm{\Sigma}_2 & \dots &  \rho_{2I}\bm{L}_2\bm{L}_I^T \\
    \vdots & \vdots & \ddots & \vdots\\
    \rho_{I1}\bm{L}_I\bm{L}_1^T & \rho_{I2}\bm{L}_I\bm{L}_2^T & \dots & \bm{\Sigma}_I
    \end{bmatrix}_{IJ\times IJ},\\ &\bm{\Sigma}_{i}=\bm{L}_i\bm{L}_i^T=\sigma_i^2\bm{D}(\bm{I}-\phi_i\bm{C})^{-1}, \rho_{ii'}=\rho_{i'i}\in [-1,1]
    \end{aligned}
\end{equation}
We name this model as a joint spatial conditional auto-regressive model and use Markov-chain Monte Carlo (MCMC) to fit the model. We give priors to the unknown parameters:  for $k\in\{0,1,...,K\}$, $\beta_{ki}$ follows a normal distribution with mean $0$ and variance $100$, denoted as $\beta_{ki}\sim\mathcal{N}(0,100)$; $\sigma^2_i$ follows a inverse gamma distribution with a shape parameter $0.1$ and rate parameter $0.1$, denoted as $1/\sigma^2_i\sim\mathcal{GA}(0.1,0.1)$; $\phi_i$ follows a uniform distribution ranging from 0 to 1, denoted as $\phi_i\sim\mathcal{U}(0,1)$; $\rho_{ii'}$ follows a a uniform distribution ranging from -1 to 1, denoted as $\rho_{ii'}\sim\mathcal{U}(-1,1)$. 

The priors of $\beta_{ki}$ and $\sigma^2_i$ are known to be conjugate priors and frequently used in Bayesian analysis of the (generalized) linear model \citep{gelman2013bayesian}, and our specification brings weak prior information. The uniform prior on $\phi_i$ has been implemented in several reports \citep[e.g.,][]{lee2013carbayes,xue2018bayesian}. The uniform prior of $\rho_{ii'}$ follows the practice of \citet{recta2012two}. We use \code{NIMBLE} \citep{de2017programming} codes to implement our proposal and the codes are attached in Section \ref{sec:codes} of the supplementary materials. 

In the above statement of this model, we treat all combinations of $(i,j,k)$ as observed ones. However, note that in our motivating data, many combinations of $(i,j,k)$ are not observed. These unobserved prevalence rates (or their transformed means) can be imputed by the predictive density function $f(\bm{\mu}^{(\mathcal{M})}|\bm{\mu}^{(\mathcal{O})},.)$, where $\bm{\mu}^{(\mathcal{M})}$ is a vector of transformed means whose indices are the missing entries and $\bm{\mu}^{(\mathcal{O})}$ is a vector of transformed means whose indices are the observed entries. The predictive density function can be intuitively understood as an extended kriging, borrowing information not only from neighbouring locations but also dependent populations. The \code{NIMBLE} codes adopt MCMC imputation \citep{de2017programming} to impute these unobserved prevalence rates for each MCMC iteration via drawing a sample from the predictive density function.

\section{Model Comparison and Evaluation}  
\label{sec:Applications}
%Given our scientific objective which is to impute the missing entries of key populations, we apply our proposal and other benchmark methods to the real HIV data described in Section \ref{sec:data} and evaluate their performances via cross-validation. The benchmark methods and our proposal are summarized in Table \ref{tab:methods} below. The table clearly outlines the distinguishing future of our proposal: our random effects capture the spatial dependence and cross-population dependence. Meanwhile, the mixed model and the CAR model only capture the spatial effects.

{Given our scientific objective, which is to impute the missing HIV prevalence among different key populations, we apply our proposal and other benchmark methods to the HIV surveillance data described in Section \ref{sec:data} and evaluate their performances via cross-validation. The model specifications of the benchmark methods and our proposal are summarized in Table \ref{tab:methods}. They are distinguished by their random effects: the simple mixed model only assumes dependence within a combination of a key population $i$ and a location $j$; the CAR model further introduces the spatial effects; our proposal captures both the spatial dependence and cross-population dependence.}

\begin{table}[ht!]
\centering
\footnotesize
\caption{The proposal and the benchmark methods.} 
%\textcolor{red}{Suggest combine Data Model into one column so that you could put $\mu_{ij}=$ in one line, and also combine Fixed Effect into one line, use $\beta_{ik}$ instead of $\beta_{ki}$, and put $\beta_{ik}$ before $k^p$}}
\label{tab:methods}
\begin{tabular}{c|ccc}
\hline\hline
\multirow{2}{*}{Method} & \multicolumn{2}{c}{Benchmark Methods} & Proposal \\
 & Mix Method & CAR Mode & Joint CAR Model \\ \hline\hline
Data Model & \multicolumn{3}{c}{$Y_{ijk}|p_{ijk}\sim \mathcal{B}(N_{ijk},p_{ijk}),\quad\mu_{ijk}=\log\frac{p_{ijk}}{1-p_{ijk}}=v_i(k)+s_{ij}$} \\ \hline
Fixed Effect & \multicolumn{3}{c}{$v_i(k)=\beta_{0i}+\sum_{p=1}^3 k^p\beta_{ki}$} \\ \hline
Random Effect & $s_{ij}\sim\mathcal{N}(0,\sigma^2_i)$ & $\bm{s}_i\sim \mathcal{N}(\bm{0},\sigma_i^2\bm{D}(\bm{I}-\phi_i\bm{C})^{-1})$ & $[\bm{s}_{1}^T, \bm{s}_{2}^T, ..., \bm{s}_{I}^T]^T\sim \mathcal{N}(\bm{0},\bm{S})$\\
\hline\hline
\end{tabular}
\end{table}

In the following numerical studies, we {randomly mark} some observed entries $Y_{ijk}$ in the real data as \textit{missing} entries. Let $\mathbb{H}$ be a set of combinations of indexes $(i,j,k)$ which are \textit{missing} and $|\mathbb{H}|$ is the size of this set. Because the true prevalence rates are unknown, the naive prevalence rate estimators {$\hat{p}_{ijk}=\frac{Y_{ijk}}{N_{ijk}}$} are used for accuracy evaluation. The accuracy is summarized in terms of mean square error, {$\textup{MSE} = \sum_{(i,j,k)\in \mathbb{H}}\frac{1}{|\mathbb{H}|}(\mathbb{E}p_{ijk}-\hat{p}_{ijk})^2$}, and $99\%$ posterior coverage on these \textit{missing} entries. Given $N$ posterior samples of $p_{ijk}$ for $(i,j,k)\in\mathbb{H}$, denoted as $\{p_{ijk}^{(t)}: t\in\{1,2,...,N\}\}$, we draw samples $\{{Y}_{ijk}^{(t)}: t\in\{1,2,...,N\}\}$ via ${Y}_{ijk}^{(t)}\sim\mathcal{B}(N_{ijk},p_{ijk}^{(t)})$, and then compute {$\{\tilde{p}_{ijk}^{(t)}=\frac{{Y}_{ijk}^{(t)}}{N_{ijk}}: t\in\{1,2,...,N\}\}$}. For $(i,j,k)\in \mathbb{H}$, we {obtain $\mathbb{E}p_{ijk}=\frac{1}{N}\sum_{t=1}^N {p}_{ijk}^{(t)}$ for MSE calculation.} We obtain the empirical $99\%$ posterior interval of the density $\{\tilde{p}_{ijk}^{(t)}=\frac{{Y}_{ijk}^{(t)}}{N_{ijk}}: t\in\{1,2,...,N\}\}$ and calculate the frequency that the empirical interval covers the naive estimator {$\hat{p}_{ijk}$}. 

To demonstrate that our proposal produces better imputations universally, we applied the methods to the HIV prevalence data of the three representative countries: Ukraine, Morocco, and Jamaica\footnote{Because Jamaica has few observations of IDUs, FSW, and MSM, only Clients and the remaining people are included in the model fitting.}. We collect $30,000$ MCMC samples discarding the first $20,000$ MCMC samples as burn-in. 

The way to partition the HIV epidemic data for cross-validation may play an important role in model evaluation \citep{gasch2015spatio,meyer2016mapping,meyer2018improving}. \citet{meyer2018improving} compared the performances of cross-validation with different partition strategies and suggested the so-called \textit{leave-one-location-out} cross-validation regarding our case, because our model treats years as replications and aims to predict unknown locations for each replication (year). The \textit{leave-one-location-out} cross-validation is described as follows. For each fold, we treat all of a key population's observations within one location over all the years as missing. {We fit the model to the rest of the observed data and calculate the MSE and the posterior coverage of the missing location based on the prediction using the posterior predictive densities.} We give the weighted average of these MSEs and posterior coverages along with their sample standard deviation (SD) in Table \ref{tab:location}. The weights for the weighted average are the missing entries of each component. The \textit{leave-one-location-out} cross-validation demonstrates that {the joint CAR model} (our proposal) produces {the best} {point estimates} and {the most} appropriate uncertainties. Both the joint CAR model and the CAR model surpass the mixed model, indicating the importance of borrowing information from neighboring locations. The difference between the performances of the {Joint CAR} and {CAR} are due to whether the cross-population dependence is taken into consideration. This motivates us to investigate the impact of the cross-population dependence on missing data imputation (Section \ref{sec:Simulation}).

% \usepackage{multirow}
\begin{table}[ht!]
\scriptsize
\centering
\caption{The results of \textit{leave-one-location-out} cross-validation.}
\label{tab:location}
\begin{tabular}{c|c|cccccc|c}
\hline\hline
\multirow{2}{*}{Index} & \multirow{2}{*}{Key Population} & \multicolumn{2}{c}{Mixed} & \multicolumn{2}{c}{CAR} & \multicolumn{2}{c|}{Joint CAR} & \multirow{2}{*}{Country} \\
 &  & Mean & SD & Mean & SD & Mean & SD &  \\ \hline\hline
\multirow{9}{*}{MSE} & FSW & 6.53E-03 & 8.61E-03 & 7.47E-03 & 9.93E-03 & \textbf{6.41E-03} & 1.03E-02 & \multirow{4}{*}{Ukraine} \\
 & MSM & \textbf{3.73E-03} & 5.65E-03 & 3.93E-03 & 7.14E-03 & 4.31E-03 & 6.59E-03 &  \\
 & IDUs & 2.19E-02 & 2.95E-02 & 2.38E-02 & 2.92E-02 & \textbf{1.62E-02} & 2.34E-02 &  \\
 & Clients & 7.57E-04 & 8.26E-04 & 7.92E-04 & 6.97E-04 & \textbf{7.51E-04} & 6.88E-04 &  \\ \cline{2-9} 
 & FSW & 4.02E-04 & 3.55E-04 & 3.41E-04 & 7.87E-04 & \textbf{2.20E-04} & 4.52E-04 & \multirow{4}{*}{Morocco} \\
 & MSM & \textbf{5.37E-04} & 5.22E-04 & 6.13E-04 & 7.65E-04 & 6.44E-04 & 9.10E-04 &  \\
 & IDUs & 8.77E-02 & 2.74E-04 & 1.01E-01 & 4.68E-03 & \textbf{8.23E-02} & 1.75E-02 &  \\
 & Clients & 6.41E-05 & 1.94E-04 & 4.99E-03 & 1.11E-02 & \textbf{1.53E-05} & 4.70E-05 &  \\ \cline{2-9} 
 & Clients & 1.93E-04 & 1.92E-04 & 1.04E-04 & 1.13E-04 & \textbf{1.12E-04} & 1.33E-04 & Jamaica \\ \hline\hline
\multirow{9}{*}{Coverage} & FSW & \textbf{98.92\%} & 3.38E-01 & 96.77\% & 3.00E-01 & 92.47\% & 2.80E-01 & \multirow{4}{*}{Ukraine} \\
 & MSM & \textbf{98.84\%} & 2.01E-01 & 91.86\% & 2.53E-01 & 95.35\% & 2.16E-01 &  \\
 & IDUs & \textbf{95.90\%} & 2.69E-01 & 95.08\% & 2.68E-01 & 94.26\% & 2.61E-01 &  \\
 & Clients & 95.00\% & 2.65E-01 & 95.00\% & 2.65E-01 & \textbf{100.00\%} & 2.42E-01 &  \\ \cline{2-9} 
 & FSW & \textbf{98.46\%} & 5.65E-01 & 95.38\% & 5.83E-01 & \textbf{98.46\%} & 6.19E-01 & \multirow{4}{*}{Morocco} \\
 & MSM & \textbf{93.02\%} & 4.56E-01 & \textbf{93.02\%} & 4.56E-01 & 81.40\% & 4.76E-01 &  \\
 & IDUs & \textbf{66.67\%} & 3.14E-01 & 55.56\% & 1.57E-01 & \textbf{66.67\%} & 3.14E-01 &  \\
 & Clients & 97.62\% & 5.41E-01 & 96.43\% & 5.15E-01 & \textbf{98.81\%} & 5.93E-01 &  \\ \cline{2-9} 
 & Clients & 82.75\% & 3.13E-01 & \textbf{84.21\%} & 3.17E-01 & \textbf{84.21\%} & 3.17E-01 & Jamaica \\ \hline\hline
\end{tabular}
\end{table}

\begin{figure}[ht!]
    \centering
    \includegraphics[width=1\textwidth]{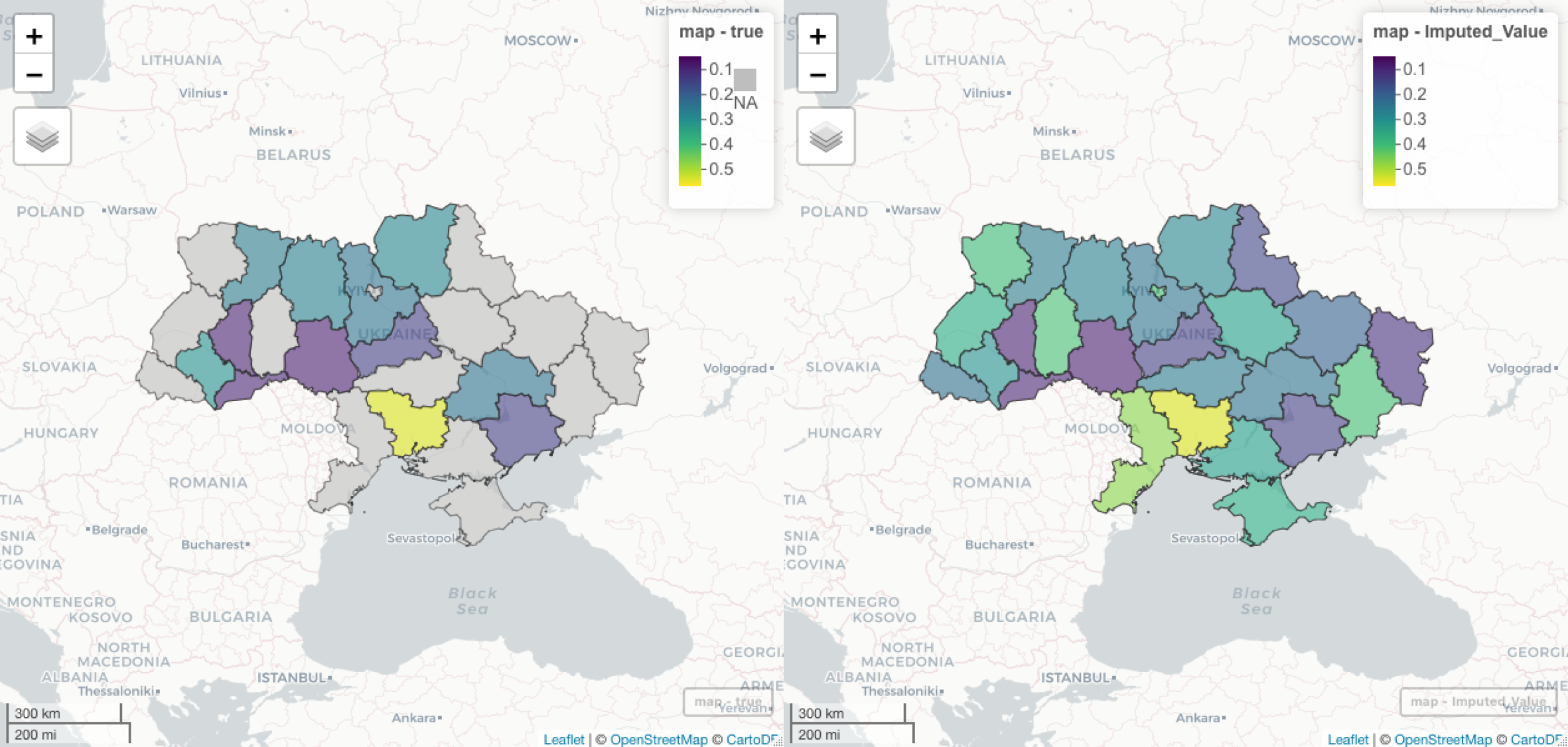}
    \caption{The left panel presents the original prevalence map where the entries are either unobserved (grey) or labeled (color) with native prevalence estimator $\hat{p}_{ijk}$. The right panel additionally use the posterior mean $\mathbb{E}{p}_{ijk}$ to impute the observed entries.}
    \label{fig:typical}
\end{figure}

We finally use the prevalence rate of the IDUs in 2009 as an illustrative example (Figure \ref{fig:typical}). We use the left panel of Figure \ref{fig:typical} to present the original prevalence map where the entries are either unobserved or labeled with native prevalence estimator $\hat{p}_{ijk}$. After model fitting, we use the posterior mean $\mathbb{E}{p}_{ijk}$ to impute the observed entries, which are presented in the right panel of Figure \ref{fig:typical}. 

\section{Impact of Missing Structure on Missing Imputation} 
\label{sec:Simulation}

\begin{figure}[ht!]
    \centering
    \includegraphics[width=0.9\textwidth]{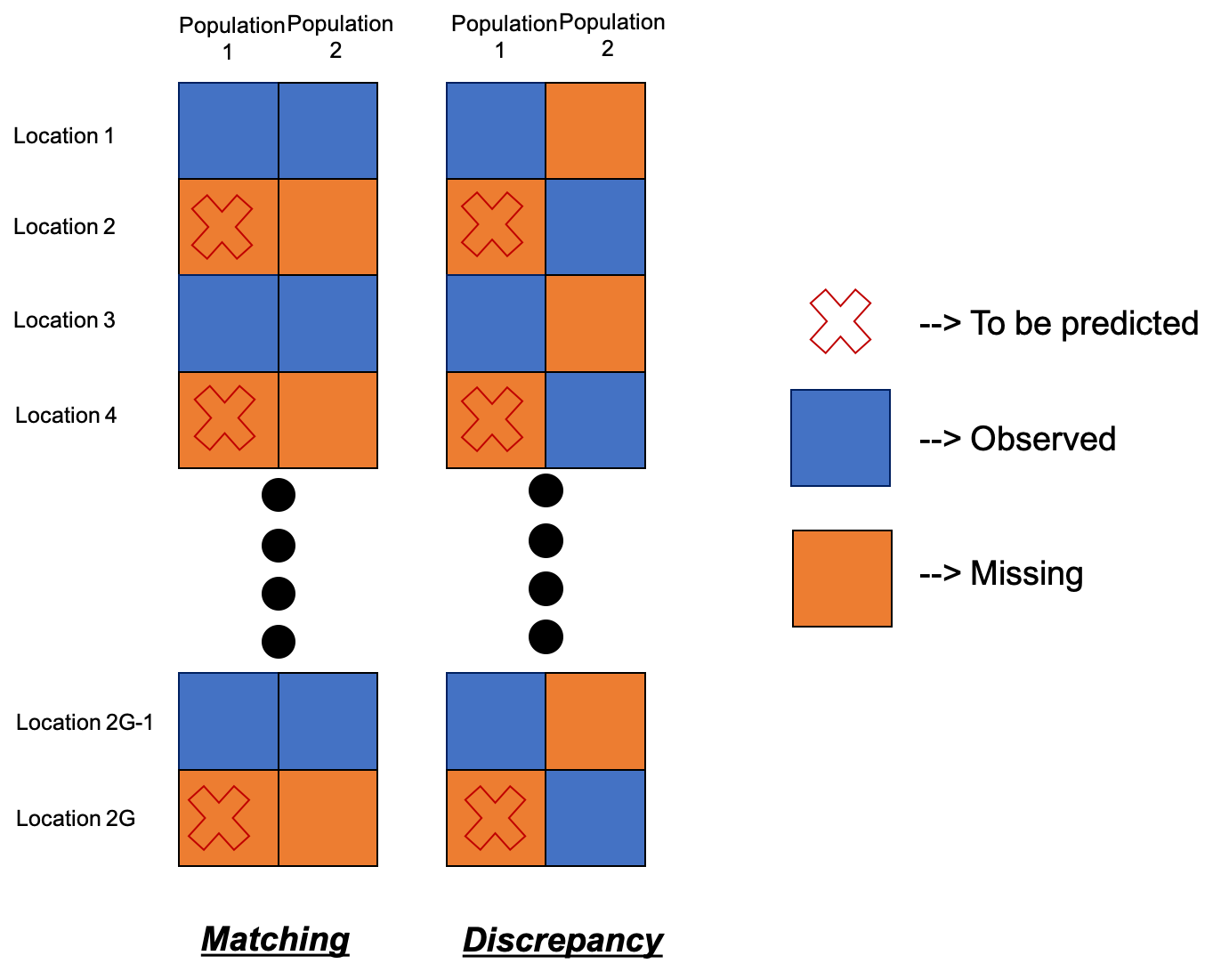}
    \caption{A graphical illustration of the two missing structures. {We assume there are $2G$ locations in total.} The blue boxes indicate the ones which are observed. The red boxes indicate the ones which are missing. The boxes labeled with a crossing indicate the ones which are to be predicted.}
    \label{fig:Missing_Pattern}
\end{figure}

{In this section, we further investigate when our proposal is expected to improve the accuracy of the missing data imputation. We conjecture that the strength of the cross-population dependence varies with the structure of the missing data.} {Considering two populations of a certain year, we define two extreme missing structures as follows:
\begin{itemize}
    \item \textit{Matching} -- at each location, the surveillance data of the two populations are either both available OR both missing;
    \item \textit{Discrepancy} -- at each location, the surveillance data are available for one population AND missing for the other population.
\end{itemize}
 A visual illustration is in Figure \ref{fig:Missing_Pattern}. The missing structure of real surveillance data for multiple populations is usually a mix of those two extremes. However, studying missing structure via focusing on the extreme cases make it efficient to evaluate their impact on missing imputation}.

In the following two sections, we study the impacts of the two missing structures on two aspects: (1) imputation robustness and (2) statistical inference of population dependence parameter $\rho_{ii'}$. The relevant derivations and proofs are summarized in Section \ref{sec:proof} of the supplementary materials.

\begin{comment}

For a concise illustration, we use the example in Figure \ref{fig:Missing_Pattern}, where where there are two populations ($I=2$), even number of locations ($J=2G$), and one year ($K=1$). For a more concise illustration, we omit the dummy index $k$ and assume that the fixed effects are zeros, such as $[\bm{\mu}_1^T, \bm{\mu}_2^T]^T=[\mu_{11}, {\mu_{12}} ,\mu_{21}, {\mu_{22}}]^T\sim\mathcal{N}\left(\bm{0},\begin{bmatrix}
    \bm{\Sigma}_1 & \rho\bm{L}_1\bm{L}_2^T \\
    \rho\bm{L}_2\bm{L}_1^T & \bm{\Sigma}_2 
    \end{bmatrix}\right)$. We aim to estimate the missing entries of population 1, denoted as $\bm{\mu}_{pred}=[\mu_{12}, \mu_{14}, ..., \mu_{1,2G}]^T$. 

\end{comment}

\subsection{Imputation Robustness}
\label{sec:imp}
In this subsection, we investigate the impact of two missing data structures {in Figure  \ref{fig:Missing_Pattern}} on imputation robustness. {For simplicity, we assume that there is only one year ($K=1$), which means that the dummy index $k$ denoting the years is omitted in the following illustration.} We evaluate the performance of imputation by treating crossed entries as {the} ones to be predicted. The missing entries are predicted by using predictive posterior density, and the densities under different missing structures are differently expressed. For the matching structure, the predictive posterior density {depends on} two components: the observations of population 1, denoted as $\bm{\mu}_{obs1}=[\mu_{11}, \mu_{13}, ..., \mu_{1,(2G-1)}]^T$, and the observations of population 2, denoted as $\bm{\mu}_{obs2}=[\mu_{21}, \mu_{23}, ..., \mu_{2,(2G-1)}]^T$.  For the discrepancy structure, the predictive posterior density {depends on} two components: the observations of population 1, denoted as $\bm{\mu}_{obs1}=[\mu_{11}, \mu_{13}, ..., \mu_{1,(2G-1)}]^T$, and the observations of population 2, denoted as $\bm{\mu}_{obs2}=[\mu_{22}, \mu_{24}, ..., \mu_{2,2G}]^T$. For both cases, we aim to predict $\bm{\mu}_{pred}=[\mu_{12}, \mu_{14}, ..., \mu_{1,2G}]^T$.  For a more concise illustration, we further assume fixed effects are zeros. Let $\bm{\Sigma}_{pred}=\bm{L}_{pred}\bm{L}_{pred}^T$, $\bm{\Sigma}_{obs1}=\bm{L}_{obs1}\bm{L}_{obs1}^T$,  $\bm{\Sigma}_{obs2}=\bm{L}_{obs2}\bm{L}_{obs2}^T$ be the marginal covariance matrices of $\bm{\mu}_{pred},\bm{\mu}_{obs1},\bm{\mu}_{obs2}$, respectively; $\rho$ be the population dependence parameter between population 1 and population 2. {We express  $\bm{\mu}_{pred},\bm{\mu}_{obs1},\bm{\mu}_{obs2}$ as}

\begin{equation*}
\begin{aligned}
    \text{Matching:}&\left\{
    \begin{aligned}
    \bm{\mu}_{pred}&=\bm{L}_{pred}(\bm{R}\bm{Z}_1+(\bm{I}-\bm{R}\bm{R}^T)^{\frac{1}{2}}\bm{Z}_2),\\
    \bm{\mu}_{obs1}&=\bm{L}_{obs1}\bm{Z}_1,\\
    \bm{\mu}_{obs2}&=\bm{L}_{obs2}(\rho\bm{Z}_1+(1-\rho^2)^{\frac{1}{2}}\bm{Z}_3),
    \end{aligned}
    \right.\\
    \\
     \text{Discrepancy:}&\left\{
    \begin{aligned}
    \bm{\mu}_{pred}&=\bm{L}_{pred}\bm{Z}_1, \\
    \bm{\mu}_{obs1}&= \bm{L}_{obs1}(\bm{R}^T\bm{Z}_1+(\bm{I}-\bm{R}^T\bm{R})^{\frac{1}{2}}\bm{Z}_2),\\
    \bm{\mu}_{obs2}&=\bm{L}_{obs2}(\rho\bm{Z}_1+(1-\rho^2)^{\frac{1}{2}}\bm{Z}_3),
    \end{aligned}
    \right.
\end{aligned}
\end{equation*}
where $\bm{R}=\bm{L}_{pred}^{-1}Cov( \bm{\mu}_{pred}, \bm{\mu}_{obs1})(\bm{L}_{obs1}^T)^{-1}$, $\bm{Z}_1$, $\bm{Z}_2$ and $\bm{Z}_3$ are independently distributed as a multivariate normal distribution with mean $\bm{0}$ and variance $\bm{I}$, and $\bm{A}^{\frac{1}{2}}$ returns the lower Cholesky factor of $\bm{A}$. Therefore, the joint {distribution} of $[\bm{\mu}_{pred},\bm{\mu}_{obs1},\bm{\mu}_{obs2}]$ {is}
\begin{equation}
    \begin{aligned}
     &\text{Matching:}&\quad \left[ \begin{bmatrix}\bm{\mu}_{pred}\\\bm{\mu}_{obs1}\\\bm{\mu}_{obs2}\end{bmatrix}|.\right]\sim\mathcal{N}\left(\bm{0}, \begin{bmatrix}\bm{\Sigma}_{pred} & \bm{L}_{pred}\bm{R}\bm{L}_{obs1}^T & \rho\bm{L}_{pred}\bm{R}\bm{L}_{obs2}^T\\
     \bm{L}_{obs1}\bm{R}^T\bm{L}_{pred}^T& \bm{\Sigma}_{obs1} & \rho\bm{L}_{obs1}\bm{L}_{obs2}^T\\
     \rho\bm{L}_{obs2}\bm{R}^T\bm{L}_{pred}^T&\rho \bm{L}_{obs2}\bm{L}_{obs1}^T &\bm{\Sigma}_{obs2}\end{bmatrix} \right)\\
     \\
        &\text{Discrepancy:}&\quad \left[ \begin{bmatrix}\bm{\mu}_{pred}\\\bm{\mu}_{obs1}\\\bm{\mu}_{obs2}\end{bmatrix}|.\right]\sim\mathcal{N}\left(\bm{0}, \begin{bmatrix}\bm{\Sigma}_{pred} & \bm{L}_{pred}\bm{R}\bm{L}_{obs1}^T & \rho\bm{L}_{pred}\bm{L}_{obs2}^T\\
     \bm{L}_{obs1}\bm{R}^T\bm{L}_{pred}^T& \bm{\Sigma}_{obs1} & \rho\bm{L}_{obs1}\bm{R}^T\bm{L}_{obs2}^T\\
     \rho\bm{L}_{obs2}\bm{L}_{pred}^T&\rho \bm{L}_{obs2}\bm{R}\bm{L}_{obs1}^T &\bm{\Sigma}_{obs2}\end{bmatrix} \right).
    \end{aligned}
\end{equation}
{Given the joint densities, we have the predictive posterior densities expressed as follows:}
%\textcolor{red}{Together with the prior distributions defined in Section \ref{sec:Method}, we obtain the following predictive distributions for $\bm{\mu}_{pred}$:}
\begin{equation}
    \begin{aligned}
    &\text{Matching:}&\quad  \\
    &[\bm{\mu}_{pred}|\bm{\mu}_{obs1},\bm{\mu}_{obs2},. ]\sim\mathcal{N}[\bm{L}_{pred}\bm{R}\bm{L}_{obs1}^T\bm{\Sigma}_{obs1}^{-1}\bm{\mu}_{obs1},\bm{\Sigma}_{pred}-\bm{L}_{pred}\bm{R}\bm{R}^T\bm{L}_{pred}^T], \\
    \\
    &\text{Discrepancy:}& \\
    &[\bm{\mu}_{pred}|\bm{\mu}_{obs1},\bm{\mu}_{obs2},.]\sim\mathcal{N}[\bm{W}_1\bm{\mu}_{obs1}+\bm{W}_2\bm{\mu}_{obs2},\bm{\Sigma}_{pred}-(\bm{W}_1\bm{L}_{obs1}\bm{R}^T\bm{L}_{pred}^T+\bm{W}_2\bm{L}_{obs1}\bm{L}_{pred}^T)],
    \end{aligned}
    \label{eq:pred}
\end{equation}
where \begin{equation}
    \begin{aligned}
    \bm{W}_1&=\bm{L}_{pred}\bm{R}(\bm{I}-\rho^2\bm{R}^T\bm{R})^{-1}\bm{L}_{obs1}^{-1}-\bm{L}_{pred}\rho^2(\bm{I}-\rho^2\bm{R}\bm{R}^T)^{-1}\bm{R}\bm{L}_{obs1}^{-1},\\
   \bm{W}_2&=-\bm{L}_{pred}\rho\bm{R}\bm{R}^T(\bm{I}-\rho^2\bm{R}\bm{R}^T)^{-1}\bm{L}_{obs2}^{-1}+\bm{L}_{pred}\rho(\bm{I}-\rho^2\bm{R}\bm{R}^T)^{-1}\bm{L}_{obs2}^{-1}.
    \end{aligned}
\end{equation} 
{In Equation (\ref{eq:pred}),} the discrepancy structure borrows information from both populations whereas the matching structure only borrows information from population 1. {Thus the Bayesian estimation under the discrepancy structure utilizes more information, which is expected to perform better than that of the matching structure.} {Taking a closer look at the predictive distributions, we find that} both predictive posterior densities enjoy unbiased mean after integrating out the observed ones, i.e., $\bm{\mu}_{obs1},\bm{\mu}_{obs2}$. However, the discrepancy structure has a smaller predictive variance {than that of the matching structure} for any $\rho\in [-1,1]$ and the difference is larger if $|\rho|$ is closer to 1. The two variances are equal if and only if $\rho=0$. Thus, $\rho$ {controls} how much information is borrowed from the dependent populations. In summary, we can give a remark below:

\begin{remark}
\label{rmk:1}
The prediction of discrepancy structure is more robust than that of the matching structure. The prediction of discrepancy structure borrows information from both populations but {the} prediction of matching structure only borrows information from neighboring locations of its own population.
\end{remark}

\begin{figure}[ht!]
    \centering
    \includegraphics[width=1\textwidth]{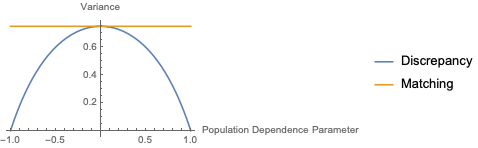}
    \caption{The relationship between the predictive variance and the population dependence parameter under two missing data structures: discrepancy in blue and matching in yellow. We assume that $G=1$ and spatial covariance matrix of both population 1 and population 2 are $\bm{\Sigma}_1=\bm{\Sigma}_2$.} %$\bm{\Sigma}_1=\bm{\Sigma}_2=\begin{bmatrix}1 & 0.5 \\ 0.5 & 1\end{bmatrix}$.
    \label{fig:Variance}
\end{figure}

The relevant proof is in the Section B.1 of the supplementary materials. Here, we use a simple example to illustrate our claim. Let $G=1$, Figure \ref{fig:Variance} illustrates the relationship between the predictive variance and the population dependence parameter under two missing data structures: discrepancy in blue and matching in yellow. {We assume that} the spatial covariance matrix of both population 1 and population 2 {are} $\bm{\Sigma}_1=\bm{\Sigma}_2=\begin{bmatrix}1 & 0.5 \\ 0.5 & 1\end{bmatrix}$. The discrepancy structure produces a smaller {predictive} variance when the absolute value of population dependence parameter is large. The population dependence parameter has no effect on the {predictive} variance of matching structure.

\subsection{Population Dependence Parameter}
\label{sec:par}
As discussed in Section \ref{sec:imp}, the discrepancy structure provides a robust prediction. However, {we also realize that} this robustness relies on a valid statistical inference of the population dependence parameter $\rho$. In this section, we discuss the impact of the missing structures on {the estimation} of the population dependence parameter $\rho$. {In particular, we compare the variances of the unbiased estimate, $\hat{\rho}=f(\bm{\mu}_{obs1},\bm{\mu}_{obs2})$, under two missing data structures.} Both $\bm{\mu}_{obs1}$ and $\bm{\mu}_{obs2}$ are observed ones as defined in Section \ref{sec:imp}. A lower bound on the variance of the unbiased estimator can be obtained by using the Cramér–Rao inequality \citep{gart1959extension}, such as $Var(\hat{\rho})\leq \mathcal{I}^{-1}(\rho)$. Given a missing structure, the inverse of the Fisher information, $\mathcal{I}^{-1}(\rho)$, is expressed as follows:

\begin{equation}
    \begin{aligned}
    &\text{Matching:}&\quad &\mathcal{I}_{Matching}^{-1}(\rho)=\frac{1}{G}\frac{(1-\rho^2)^2}{1+\rho^2}.\\
    &\text{Discrepancy:}&\quad &\mathcal{I}_{Discrepancy}^{-1}(\rho)=1/Tr[\rho\bm{R}^T(\bm{I}-\rho^2\bm{R}\bm{R}^T)^{-1}\bm{R}]^2.
    \end{aligned}
\end{equation}

%\textcolor{gray}{We find that $\mathcal{I}_{Matching}^{-1}(\rho)<\mathcal{I}_{Discrepancy}^{-1}(\rho)$ for any $\rho$ and $\bm{R}$. If the estimators satisfy that $a(\rho)(\hat{\rho}-\rho)=\frac{d}{d\rho}\log L(\rho|\bm{\mu}_{obs1},\bm{\mu}_{obs2})$ where $a(\rho)$ is a function of $\rho$ and $\log L(\rho|\bm{\mu}_{obs1},\bm{\mu}_{obs2})$ is the likelihood log of $[\bm{\mu}_{obs1},\bm{\mu}_{obs2}]$ with ${Y}_{ijk}$ marginalized, the lower bands are attained and the estimator of matching structure is more efficient. Furthermore, if $\bm{R}=\bm{0}$ where there is no spatial dependence, the lower bond is infinity. In summary, we can give a remark below:}

{The above information bounds are attained if $\frac{d}{d\rho}\log L(\rho|\bm{\mu}_{obs1},\bm{\mu}_{obs2})=a(\rho)(\hat{\rho}-\rho)$ where $a(\rho)$ is a function of $\rho$ and $\log L(\rho|\bm{\mu}_{obs1},\bm{\mu}_{obs2})$ is the log likelihood of $[\bm{\mu}_{obs1},\bm{\mu}_{obs2}]$ with ${Y}_{ijk}$ marginalized. In addition, $\mathcal{I}_{Matching}^{-1}(\rho)<\mathcal{I}_{Discrepancy}^{-1}(\rho)$ for any $\rho$ and $\bm{R}$ (See the proof in Supplement B.2). When $\bm{R}=\bm{0}$, $\mathcal{I}_{Discrepancy}^{-1}(\rho)$ is infinity meaning that $\rho$ could not be estimated. In summary, we can give a remark below:}

\begin{remark}
\label{rmk:2}
The estimator of the population dependence parameter under the matching structure is more efficient than that under the discrepancy structure.
\end{remark}

\begin{figure}[ht!]
    \centering
    \includegraphics[width=0.6\textwidth]{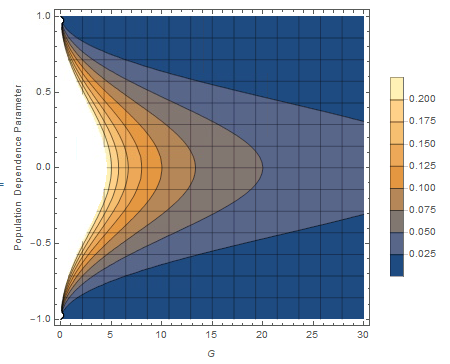}
    \caption{The contour plot of $\mathcal{I}_{Matching}^{-1}(\rho)$. The x-axis is for number of pairs. The y-axis is for the population dependence parameter.}
    \label{fig:densityplot}
\end{figure}

{Figure \ref{fig:densityplot} shows how $\mathcal{I}_{Matching}^{-1}(\rho)$ varies by $\rho$ and $G$, which provides a way to control the variance of $\hat{\rho}$ in practice.} {For instance, if $|\rho|\geq 0.5$, then observing data for both populations in $18$ locations could ensure $\mathcal{I}_{Matching}^{-1}(\rho)\leq0.025$.}
If we take $K$ years/replications into consideration, the inverted fisher information is $\frac{1}{G\times K}\frac{(1-\rho^2)^2}{1+\rho^2}$. This means the {needed matching locations} can be accumulative over the years. {The examples we considered in this article satisfied} this requirement.

\begin{figure}[ht!]
    \centering
    \includegraphics[width=1\textwidth]{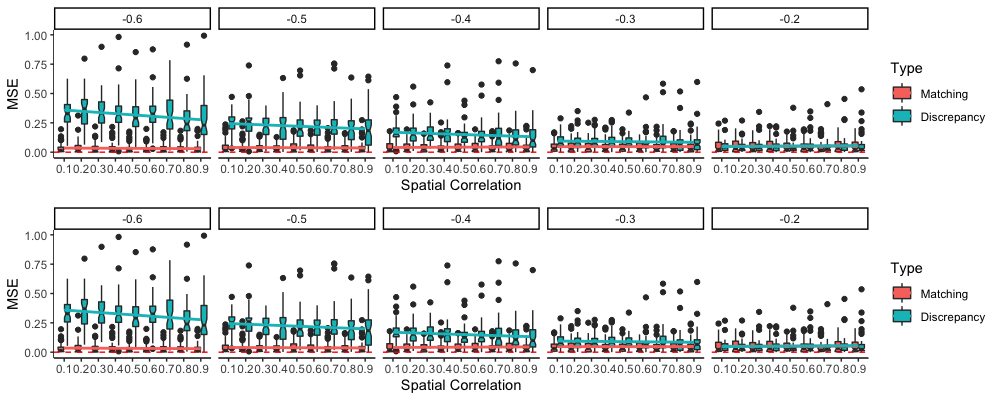}
    \caption{The MSEs of the population dependence parameter varied by spatial correlation parameter $\phi_{1}=\phi_2$ (x-axis) and population dependence parameter $\rho_{12}$ (value on the top of each figure).}
    \label{fig:PE1}
\end{figure}

To further validate Remark \ref{rmk:2}, we also numerically investigate how the spatial parameters affect the statistical inference of $\rho_{ii'}$. The simulated data are generated based on our proposed model (Equation \ref{eq:proposed}). We use the map of Ukraine. We have $I=2$, $J=27$, and $K=10$. For all $i,j,k$, we have the sample size $N_{ijk}=100$. We give the population-specific trends as $v_1(k)=\sin(0.2k)$ and $v_2(k)=\cos(0.2k)$. The population-specific random effects are generated as $\bm{s}_i\sim\mathcal{N}(\bm{0},\bm{S})$. The parameters associated with $\bm{S}$ are specified as follows. The population-specific spatial parameters are specified as $\sigma_1^2=\sigma_2^2=1$. For each run, we sample a simulated data as follows, we simulate $50$ replications with $\rho_{12}\in\{0.1, 0.2, ..., 0.9\}$ and $\phi_{1}=\phi_2\in\{0.2,0.3,...,0.6,0.7\}$ and hide the assumed missing entries as presented in Figure \ref{fig:Missing_Pattern}. From Figure \ref{fig:PE1}, we can find that the MSE of posterior mean of $\rho$ is large under the structure of discrepancy. The MSE increases if the absolute value of $\rho$ is large. However, the structure of matching does not impact the posterior mean of $\rho_{12}$. All these are consistent with our previous claims using Cramér–Rao inequality \citep{gart1959extension}.

\subsection{Missing Structure Trade-Off}
Putting Remarks \ref{rmk:1} and \ref{rmk:2} together, we found a trade-off between two missing structures: \textit{The matching structure gives an efficient estimation of population dependence parameter, but its prediction {does not use the} cross-population information} v.s. \textit{The discrepancy structure gives a robust prediction by borrowing cross-population information but leads to inefficient estimation of the population dependence parameter}. Given that the cross-population dependence induced by the \textit{Gaussian cosimulation} \citep{oliver2003gaussian} has a variety of applications environment \citep[e.g.,][]{recta2012two,fanshawe2012bivariate}, this trade-off helps understand the strength and limitation of this model in terms of missing data imputation.

\section{Conclusion and Discussion}
\label{sec:Conlusion}

%\textcolor{gray}{Obtaining accurate HIV prevalence rates is an key bottleneck in HIV prevention. In this paper, we propose a generalized polynomial regression mixing with a joint conditional auto-regressive (CAR) model, which fully utilizes existing data. The fully utilization of existing data means more possible variations, i.e., spatial variation and cross-population dependence, are captured by our model. A substantial improvement in data imputation is obtained in the real data application, primarily resolving our scientific goal of estimating the HIV prevalence rates among key populations. This provides a useful statistical tool for HIV epidemiologists to impute the unknown prevalence rates and reveal potential spatial/cross-population variation. The study also motivates us to explore the impact of missing data on imputation accuracy. In light of this, we give a relevant investigation to reveal the strength and limitation of \textit{Gaussian cosimulation} when the model is applied to the missing data imputation. Putting all the contributions together, our paper provides both a high-impacted work of application as well as a theoretical investigation of model performance.}

{Understanding HIV prevalence among key populations is important for HIV prevention. However, accurate estimates have been difficult to obtain because their HIV surveillance data is very limited. In this paper, we propose a generalized liner mixed model with the spatial conditional auto-regressive feature which captures both the spatial dependence and the cross-population dependence. The proposed model fully utilizes existing data and provides a useful statistical tool for HIV epidemiologists to impute the unknown prevalence rates and reveal potential spatial/cross-population variation. A substantial improvement in data imputation is obtained in the real data application, primarily resolving our scientific goal of estimating the HIV prevalence rates among key populations. The study also motivates us to explore the impact of missing data on imputation accuracy. We present both simulation results and theoretical results to reveal the strength and limitation of \textit{Gaussian cosimulation} when the model is applied to the missing data imputation.}

Two topics are worthwhile for further investigation. The first {one} is the cross-population dependence. Our proposal adopts \textit{Gaussian cosimulation} \citep{oliver2003gaussian}, and the correlation parameter describes a linear relationship between the processes of any two populations. However, the actual cross-population dependence may be more complicated than the one we have proposed. Modern statistical methods such as graphical models may be implemented for handling the cross-population dependence, but the limited availability of our surveillance data may be a hurdle. The second one is the data missing mechanism. The missing entries are not missing completely at random but due to some specific reasons (e.g., resource allocation, administrative issues). It would be useful to know why the HIV surveillance data were missing for some combinations of key population, year and location, so that the potential bias due to missing not at random could be addressed. Unfortunately, such information is not readily available.

%\textcolor{gray}{Given many epidemiological reports, the missing entries are not missing completely at random but due to some specific reasons (e.g., resource allocation, administrative issues). The current missing data theories and methodologies are primarily built under the applications of clinical studies. This means the missing mechanism is usually classified based on the characteristics of a patient, e.g., missing not at random (MNAR) indicates that a patient's dropout is related to the information that is not observed. However, how to correctly specify the corresponding missing mechanisms for missing data may be a burden in applying to our motivating data, and it requires a further investigation.}

\newpage
\section*{Supplementary Materials}
\addcontentsline{toc}{section}{Appendices}
\renewcommand{\thesubsection}{\Alph{subsection}}

\subsection{Codes}
\label{sec:codes}
We compile our codes into an \code{R} package \code{JointSpCAR}. The package provides a function \code{JointCAR()} to implement our proposal. We also provide an \code{R} markdown script \code{implantation.rmd} introducing the function implementation by using synthetic HIV epidemic data. In addition, we also give the functions which are \code{CAR()} and \code{Mixed()}. They implement the benchmark methods which are the CAR model and Mixed model, respectively.

\subsection{Proofs}
\label{sec:proof}
In this section, we give essential proofs of this article. 

\subsubsection{Proofs for Section \ref{sec:imp} Imputation Robustness}
\label{sec:proof1}
In this subsection, we give the proofs for Section \ref{sec:imp} Imputation Robustness. We give that $\bm{S}_{11}=Var(\bm{\mu}_{pred})$, $\bm{S}_{22}=Var([\bm{\mu}_{obs1}^T,\bm{\mu}_{obs2}^T]^T)$, and $\bm{S}_{12}=Cov(\bm{\mu}_{pred},[\bm{\mu}_{obs1}^T,\bm{\mu}_{obs2}^T]^T)$. Thus, for both structures, the posterior mean is $\bm{S}_{12}\bm{S}_{22}^{-1}[\bm{\mu}_{obs1}^T,\bm{\mu}_{obs2}^T]^T$ and the posterior variance matrix is $\bm{S}_{11}-\bm{S}_{12}\bm{S}_{22}^{-1}\bm{S}_{12}^T$.

The bottleneck of obtaining the mean and the variance is $\bm{S}_{22}^{-1}$. \citet{lu2002inverses} gives an explicit inverse formula for a $2\times 2$ block matrix, which is summarized in Theorem \ref{thm:inverse}:
\begin{theorem}
\label{thm:inverse}
Let $\bm{R}=\begin{bmatrix}\bm{A} & \bm{B}\\ \bm{C} & \bm{D}\end{bmatrix}$ be a square positive semi-definite matrix where $\bm{A}$ and $\bm{D}$ are all square matrices. Let $\bm{R}^{-1}=\begin{bmatrix}\bm{E} & \bm{F}\\ \bm{G} & \bm{H}\end{bmatrix}$ be inversion of $\bm{R}$ and the components in the inversion are
\begin{itemize}
    \item $\bm{E}=\bm{A}^{-1}+\bm{A}^{-1}\bm{B}(\bm{D}-\bm{C}\bm{A}^{-1}\bm{B})^{-1}\bm{C}\bm{A}^{-1}$
    \item $\bm{F}=-\bm{A}^{-1}\bm{B}(\bm{D}-\bm{C}\bm{A}^{-1}\bm{B})^{-1}$
    \item $\bm{G}=-(\bm{D}-\bm{C}\bm{A}^{-1}\bm{B})\bm{C}\bm{A}^{-1}$
    \item $\bm{H}=(\bm{D}-\bm{C}\bm{A}^{-1}\bm{B})^{-1}$
\end{itemize}
\end{theorem}

Next, we apply Theorem \ref{thm:inverse} to both structures:
\paragraph{Matching Structure:} Here, we give that $\bm{A}=\bm{\Sigma}_{obs1}$, $\bm{B}=\rho\bm{L}_{obs1}\bm{L}_{obs2}^T$, $\bm{C}=\rho\bm{L}_{obs2}\bm{L}_{obs1}^T$, and $\bm{D}=\bm{\Sigma}_{obs2}$. Then, $\bm{E}$, $\bm{F}$, $\bm{G}$, and $\bm{H}$ are expressed as follows:
\begin{equation*}
    \begin{aligned}
    \bm{E}&=\bm{A}^{-1}+\bm{A}^{-1}\bm{B}(\bm{D}-\bm{C}\bm{A}^{-1}\bm{B})^{-1}\bm{C}\bm{A}^{-1}\\
    &=\bm{\Sigma}_{obs1}^{-1}+\bm{\Sigma}_{obs1}^{-1}\rho\bm{L}_{obs1}\bm{L}_{obs2}^T(\bm{\Sigma}_{obs2}-\rho\bm{L}_{obs2}\bm{L}_{obs1}^T\bm{\Sigma}_{obs1}^{-1}\rho\bm{L}_{obs1}\bm{L}_{obs2}^T)^{-1}\rho\bm{L}_{obs2}\bm{L}_{obs1}^T\bm{\Sigma}_{obs1}^{-1}\\
    &=\frac{1}{1-\rho^2}\bm{\Sigma}_{obs1}^{-1}\\
    \\
       \bm{F}&=-\bm{A}^{-1}\bm{B}(\bm{D}-\bm{C}\bm{A}^{-1}\bm{B})^{-1}\\
    &=-\bm{\Sigma}_{obs1}^{-1}\rho\bm{L}_{obs1}\bm{L}_{obs2}^T(\bm{\Sigma}_{obs2}-\rho\bm{L}_{obs2}\bm{L}_{obs1}^T\bm{\Sigma}_{obs1}^{-1}\rho\bm{L}_{obs1}\bm{L}_{obs2}^T)^{-1}\\
    &=-\bm{\Sigma}_{obs1}^{-1}\bm{L}_{obs1}\bm{L}_{obs2}^T\bm{\Sigma}_{obs2}^{-1}\frac{\rho}{1-\rho^2}\\
    \\
    \bm{G}&=-\bm{\Sigma}_{obs2}^{-1}\bm{L}_{obs2}\bm{L}_{obs1}^T\bm{\Sigma}_{obs1}^{-1}\frac{\rho}{1-\rho^2}\\
    \\
    \bm{H}&=(\bm{D}-\bm{C}\bm{A}^{-1}\bm{B})^{-1}\\
    &=(\bm{\Sigma}_{obs2}-\rho\bm{L}_{obs2}\bm{L}_{obs1}^T\bm{\Sigma}_{obs1}^{-1}\rho\bm{L}_{obs1}\bm{L}_{obs2}^T)^{-1}\\
    &=\frac{1}{1-\rho^2}\bm{\Sigma}_{obs2}^{-1}
    \end{aligned}
\end{equation*}
Then the posterior mean is 
\begin{equation*}
    \begin{aligned}
    &\bm{S}_{12}\bm{S}_{22}^{-1}[\bm{\mu}_{obs1}^T,\bm{\mu}_{obs2}^T]^T\\
    &=[\bm{L}_{pred}\bm{R}\bm{L}_{obs1}^T,\ \rho\bm{L}_{pred}\bm{R}\bm{L}_{obs2}^T]\begin{bmatrix}\frac{1}{1-\rho^2}\bm{\Sigma}_{obs1}^{-1} & -\bm{\Sigma}_{obs1}^{-1}\bm{L}_{obs1}\bm{L}_{obs2}^T\bm{\Sigma}_{obs2}^{-1}\frac{\rho}{1-\rho^2}\\ -\bm{\Sigma}_{obs2}^{-1}\bm{L}_{obs2}\bm{L}_{obs1}^T\bm{\Sigma}_{obs1}^{-1}\frac{\rho}{1-\rho^2} & \frac{1}{1-\rho^2}\bm{\Sigma}_{obs2}^{-1}\end{bmatrix}\begin{bmatrix}\bm{\mu}_{obs1}\\\bm{\mu}_{obs2}\end{bmatrix}\\
    &=\bm{L}_{pred}\bm{R}\bm{L}_{obs1}^T\bm{\Sigma}_{obs1}^{-1}\bm{\mu}_{obs1},
    \end{aligned} 
\end{equation*}
and the posterior variance is
\begin{equation*}
\small
    \begin{aligned}
    &\bm{S}_{11}-\bm{S}_{12}\bm{S}_{22}^{-1}\bm{S}_{12}^T\\
    &=\bm{\Sigma}_{pred}-\\
    &[\bm{L}_{pred}\bm{R}\bm{L}_{obs1}^T,\ \rho\bm{L}_{pred}\bm{R}\bm{L}_{obs2}^T]\begin{bmatrix}\frac{1}{1-\rho^2}\bm{\Sigma}_{obs1}^{-1} & -\bm{\Sigma}_{obs1}^{-1}\bm{L}_{obs1}\bm{L}_{obs2}^T\bm{\Sigma}_{obs2}^{-1}\frac{\rho}{1-\rho^2}\\ -\bm{\Sigma}_{obs2}^{-1}\bm{L}_{obs2}\bm{L}_{obs1}^T\bm{\Sigma}_{obs1}^{-1}\frac{\rho}{1-\rho^2} & \frac{1}{1-\rho^2}\bm{\Sigma}_{obs2}^{-1}\end{bmatrix}\begin{bmatrix}\bm{L}_{obs1}\bm{R}^T\bm{L}_{pred}^T\\\rho\bm{L}_{obs2}\bm{R}^T\bm{L}_{pred}^T\end{bmatrix}\\
    &=\bm{\Sigma}_{pred}-[\bm{L}_{pred}\bm{R}\bm{L}_{obs1}^T\bm{\Sigma}_{obs1}^{-1},\bm{0}]\begin{bmatrix}\bm{L}_{obs1}\bm{R}^T\bm{L}_{pred}^T\\\rho\bm{L}_{obs2}\bm{R}^T\bm{L}_{pred}^T\end{bmatrix}\\
    &=\bm{\Sigma}_{pred}-\bm{L}_{pred}\bm{R}\bm{R}^T\bm{L}_{pred}^T
    \end{aligned} 
\end{equation*}

\paragraph{Discrepancy Structure:} Here, we give that $\bm{A}=\bm{\Sigma}_{obs1}$, $\bm{B}=\rho\bm{L}_{obs1}\bm{R}^T\bm{L}_{obs2}^T$, $\bm{C}=\rho\bm{L}_{obs2}\bm{R}\bm{L}_{obs1}^T$, and $\bm{D}=\bm{\Sigma}_{obs2}$. Then, $\bm{E}$, $\bm{F}$, $\bm{G}$, and $\bm{H}$ are expressed as follows:
\begin{equation*}
    \begin{aligned}
    \bm{E}&=\bm{A}^{-1}+\bm{A}^{-1}\bm{B}(\bm{D}-\bm{C}\bm{A}^{-1}\bm{B})^{-1}\bm{C}\bm{A}^{-1}\\
   &=\bm{\Sigma}_{obs1}^{-1}+\bm{\Sigma}_{obs1}^{-1}\rho\bm{L}_{obs1}\bm{R}^T\bm{L}_{obs2}^T(\bm{\Sigma}_{obs2}-\rho\bm{L}_{obs2}\bm{R}\bm{L}_{obs1}^T\bm{\Sigma}_{obs1}^{-1}\rho\bm{L}_{obs1}\bm{R}\bm{L}_{obs2}^T)^{-1}\rho\bm{L}_{obs2}\bm{R}\bm{L}_{obs1}^T\bm{\Sigma}_{obs1}^{-1}\\
   &=(\bm{\Sigma}_{obs1}-\rho^2\bm{L}_{obs1}\bm{R}^T\bm{R}\bm{L}_{obs1}^T)^{-1}\\
   \\
   \bm{F}&=-\bm{A}^{-1}\bm{B}(\bm{D}-\bm{C}\bm{A}^{-1}\bm{B})^{-1}\\
   &=-\bm{\Sigma}_{obs1}^{-1}\rho\bm{L}_{obs1}\bm{R}^T\bm{L}_{obs2}^T(\bm{\Sigma}_{obs2}-\rho\bm{L}_{obs2}\bm{R}\bm{L}_{obs1}^T\bm{\Sigma}_{obs1}^{-1}\rho\bm{L}_{obs1}\bm{R}^T\bm{L}_{obs2}^T)^{-1}\\
   &=-\bm{\Sigma}_{obs1}^{-1}\rho\bm{L}_{obs1}\bm{R}^T\bm{L}_{obs2}^T(\bm{\Sigma}_{obs2}-\rho^2\bm{L}_{obs2}\bm{R}\bm{R}^T\bm{L}_{obs2}^T)^{-1}\\
   &= -\rho(\bm{L}_{obs1}^T)^{-1}\bm{R}^T(\bm{I}-\rho^2\bm{R}\bm{R}^T)^{-1}\bm{L}_{obs2}^{-1}\\
   \\
   \bm{G}&=-\rho(\bm{L}_{obs2}^T)^{-1}(\bm{I}-\rho^2\bm{R}\bm{R}^T)^{-1}\bm{R}\bm{L}_{obs1}^{-1}\\
   \\
   \bm{H}&=(\bm{D}-\bm{C}\bm{A}^{-1}\bm{B})^{-1}\\
   &=(\bm{\Sigma}_{obs2}-\rho\bm{L}_{obs2}\bm{R}\bm{L}_{obs1}^T\bm{\Sigma}_{obs1}^{-1}\rho\bm{L}_{obs1}\bm{R}^T\bm{L}_{obs2}^T)^{-1}\\
   &=(\bm{\Sigma}_{obs2}-\rho^2\bm{L}_{obs2}\bm{R}\bm{R}^T\bm{L}_{obs2}^T)^{-1}
    \end{aligned}
\end{equation*}
Then the posterior mean is 
\begin{equation*}
    \begin{aligned}
    &\bm{S}_{12}\bm{S}_{22}^{-1}[\bm{\mu}_{obs1}^T,\bm{\mu}_{obs2}^T]^T\\
    &=[\bm{L}_{pred}\bm{R}\bm{L}_{obs1}^T,\ \rho\bm{L}_{pred}\bm{L}_{obs2}^T]\\
    &\begin{bmatrix} (\bm{\Sigma}_{obs1}-\rho^2\bm{L}_{obs1}\bm{R}^T\bm{R}\bm{L}_{obs1}^T)^{-1} &  -\rho(\bm{L}_{obs1}^T)^{-1}\bm{R}^T(\bm{I}-\rho^2\bm{R}\bm{R}^T)^{-1}\bm{L}_{obs2}^{-1}\\ -\rho(\bm{L}_{obs2}^T)^{-1}(\bm{I}-\rho^2\bm{R}\bm{R}^T)^{-1}\bm{R}\bm{L}_{obs1}^{-1} & (\bm{\Sigma}_{obs2}-\rho^2\bm{L}_{obs2}\bm{R}\bm{R}^T\bm{L}_{obs2}^T)^{-1} \end{bmatrix}\begin{bmatrix}\bm{\mu}_{obs1}\\\bm{\mu}_{obs2}\end{bmatrix}\\
    &=\bm{W}_1\bm{\mu}_{obs1}+\bm{W}_2\bm{\mu}_{obs2},
    \end{aligned} 
\end{equation*}
where \begin{equation*}
    \begin{aligned}
   \bm{W}_1&=\bm{L}_{pred}\bm{R}(\bm{I}-\rho^2\bm{R}^T\bm{R})^{-1}\bm{L}_{obs1}^{-1}-\bm{L}_{pred}\rho^2(\bm{I}-\rho^2\bm{R}\bm{R}^T)^{-1}\bm{R}\bm{L}_{obs1}^{-1}\\
   \bm{W}_2&=-\bm{L}_{pred}\rho\bm{R}\bm{R}^T(\bm{I}-\rho^2\bm{R}\bm{R}^T)^{-1}\bm{L}_{obs2}^{-1}+\bm{L}_{pred}\rho(\bm{I}-\rho^2\bm{R}\bm{R}^T)^{-1}\bm{L}_{obs2}^{-1}
    \end{aligned}
\end{equation*} 
and the posterior variance is
\begin{equation*}
\small
    \begin{aligned}
    &\bm{S}_{11}-\bm{S}_{12}\bm{S}_{22}^{-1}\bm{S}_{12}^T\\
    &=\bm{\Sigma}_{pred}-[\bm{W}_1\ \bm{W}_2]\begin{bmatrix}\bm{L}_{obs1}\bm{R}^T\bm{L}_{pred}^T\\\rho\bm{L}_{obs2}\bm{L}_{pred}^T\end{bmatrix}\\
    &=\bm{\Sigma}_{pred}-\\
    &[\bm{L}_{pred}\bm{R}(\bm{I}-\rho^2\bm{R}^T\bm{R})^{-1}\bm{R}^T\bm{L}_{pred}^T-\bm{L}_{pred}\rho^2(\bm{I}-\rho^2\bm{R}\bm{R}^T)^{-1}\bm{R}\bm{R}^T\bm{L}_{pred}^T-\\
    &\bm{L}_{pred}\rho\bm{R}\bm{R}^T(\bm{I}-\rho^2\bm{R}\bm{R}^T)^{-1}\rho\bm{L}_{pred}^T+\bm{L}_{pred}\rho(\bm{I}-\rho^2\bm{R}\bm{R}^T)^{-1}\rho\bm{L}_{pred}^T]
    \end{aligned} 
\end{equation*}

Next, we want to prove that the predictive variance of the matching structure is larger than that of discrepancy structure, that is
\begin{equation*}
\small
    \begin{aligned}
    \Delta\bm{S}=\bm{S}_{Matching}-\bm{S}_{Discrepancy}\succeq \bm{0},
    \end{aligned}
\end{equation*}
where $\bm{S}_{Matching}$ and $\bm{S}_{Discrepancy}$ are the conditional covariance matrices of the matching structure and discrepancy structure, respectively. This is also equivalent to show that
%There are several means to prove that a matrix is positive semi-definite. Here, we use the fact that the determinant of every leading principal sub-matrices of $\Delta\bm{S}$ is positive if and only if $\Delta\bm{S}$ is positive semi-definite. The $k$-th principal sub-matrices of $\bm{S}_{Matching}$ and $\bm{S}_{Discrepancy}$ are the conditional variance matrices for the first $k$ components of ${\bm{\mu}_{pred}}$. That means w The $k$-th principal sub-matrices of $\bm{S}_{Matching}$ and $\bm{S}_{Discrepancy}$ has the same expression but $\bm{R}$ and ${\bm{L}_{pred}}$ are replaced by ${\bm{L}_{pred}}_{(k)}$ and $\bm{R}_{(k)}$, respectively. $\bm{R}_{(k)}={\bm{L}_{pred}}_{(k)}^{-1}Cov( {\bm{\mu}_{pred}}_{(k)}, \bm{\mu}_{obs1})({\bm{L}_{obs1}}_{(k)}^T)^{-1}$ where ${\bm{\mu}_{pred}}_{(k)}$ is the first $k$ components of ${\bm{\mu}_{pred}}$ and ${\bm{L}_{pred}}_{(k)}$ is the lower Cholesiki factor of the covariance matrix of ${\bm{\mu}_{pred}}_{(k)}$. This means, to show that the determinant of every leading principal sub-matrices of $\Delta\bm{S}$ is positive is equivalent to show the determinant of $\Delta\bm{S}$ is positive. This is also equivalent to show that
\begin{equation*}
\small
    \begin{aligned}
    &\bm{U}=\\
    &[{\bm{R}(\bm{I}-\rho^2\bm{R}^T\bm{R})^{-1}\bm{R}^T}-{\rho^2(\bm{I}-\rho^2\bm{R}\bm{R}^T)^{-1}\bm{R}\bm{R}^T}-{\rho\bm{R}\bm{R}^T(\bm{I}-\rho^2\bm{R}\bm{R}^T)^{-1}\rho}+{\rho(\bm{I}-\rho^2\bm{R}\bm{R}^T)^{-1}\rho}]-{\bm{R}\bm{R}^T}\\
    &\succeq \bm{0}
    \end{aligned}
\end{equation*}
for all $\rho\in [-1,1]$ and $\bm{R}$. Given the Neumann series, $\bm{U}$ can be expressed as

\begin{equation}
    \begin{aligned}
    \bm{U}&=\bm{R}\sum_{k=0}^{\infty}(\rho^2\bm{R}^T\bm{R})^k\bm{R}^T-\rho^2\sum_{k=0}^{\infty}(\rho^2\bm{R}\bm{R}^T)^k\bm{R}\bm{R}^T-\rho^2\bm{R}\bm{R}^T\sum_{k=0}^{\infty}(\rho^2\bm{R}\bm{R}^T)^k+\rho^2\sum_{k=0}^{\infty}(\rho^2\bm{R}\bm{R}^T)^k-\bm{R}\bm{R}^T\\
    &=\bm{R}\sum_{k=0}^{\infty}(\rho^2\bm{R}^T\bm{R})^k\bm{R}^T-2\rho^2\sum_{k=0}^{\infty}(\rho^2)^k(\bm{R}\bm{R}^T)^{k+1}+\rho^2\sum_{k=0}^{\infty}(\rho^2\bm{R}\bm{R}^T)^k-\bm{R}\bm{R}^T\\
    &=\sum_{k=1}^{\infty}(\rho^2)^k(\bm{R}\bm{R}^T)^{k+1}-2\rho^2\sum_{k=0}^{\infty}(\rho^2)^k(\bm{R}\bm{R}^T)^{k+1}+\rho^2\sum_{k=0}^{\infty}(\rho^2\bm{R}\bm{R}^T)^k\\
    &=\sum_{k=1}^{\infty}(\rho^2)^k(\bm{R}\bm{R}^T)^{k+1}-2\rho^2\sum_{k=1}^{\infty}(\rho^2)^{k-1}(\bm{R}\bm{R}^T)^{k}+\rho^2\sum_{k=1}^{\infty}(\rho^2\bm{R}\bm{R}^T)^{k-1}\\
    &=\sum_{k=1}^{\infty}\rho^{2k}\bm{Z}^{k-1}(\bm{I}-\bm{Z})^2,
    \end{aligned}
\end{equation}
where $\bm{Z}=\bm{R}\bm{R}^T$. Because $\bm{Z}^{k-1}(\bm{I}-\bm{Z})^2$ is positive semi-definite for all $k$, then $\bm{U}$ is positive semi-definite. In summary, we proved our statement.

\subsubsection{Proofs for Section \ref{sec:par} Population Dependence Parameter}
In this subsection, we give the proofs for Section \ref{sec:par} Population Dependence Parameter. The Fisher information $\mathcal{I}(\rho)$ \citep{malago2015information} is 
\begin{equation}
    \begin{aligned}
    \mathcal{I}(\rho)=\frac{1}{2}Tr(\bm{S}_{22}^{-1}\frac{\partial \bm{S}_{22}}{\partial\rho}\bm{S}_{22}^{-1}\frac{\partial \bm{S}_{22}}{\partial\rho}).
    \end{aligned}
\end{equation}
We have already given the expressions of $\bm{S}_{22}^{-1}$ and $\bm{S}_{22}$ in Section \ref{sec:proof1}. Thus, it is not difficult to give explicit expression of these $\mathcal{I}(\rho)$ by plugging in the expressions of $\bm{S}_{22}^{-1}$ and $\bm{S}_{22}$ under matching structure or discrepancy structure.

Next, we want to prove $\mathcal{I}_{Matching}^{-1}(\rho)<\mathcal{I}_{Discrepancy}^{-1}(\rho)$ for any $\rho$ and $\bm{R}$. Our goal is to find a lower bond of $\mathcal{I}_{Discrepancy}^{-1}(\rho)$. Given the Neumann series, $\mathcal{I}_{Discrepancy}^{-1}(\rho)$ can be expressed as 

\begin{equation*}
    \begin{aligned}
    \mathcal{I}_{Discrepancy}^{-1}(\rho)&=1/Tr[\rho\bm{R}^T(\bm{I}-\rho^2\bm{R}\bm{R}^T)^{-1}\bm{R}]^2\\
    &=1/Tr[\rho\bm{R}\sum_{k=0}^{\infty}(\rho^2\bm{R}^T\bm{R})^k\bm{R}^T]^2\\
    &=1/Tr[\rho\sum_{k=0}^{\infty}(\rho^2)^{k}(\bm{R}\bm{R}^T)^{k+1}]^2\\
    &=1/Tr\left[\rho^2\left(\sum_{k=0}^{\infty}(\rho^2)^{2k}(\bm{R}\bm{R}^T)^{2(k+1)}+\sum_{i\not =j}^{0:\infty}(\rho^2)^{i}(\bm{R}\bm{R}^T)^{i+1}(\rho^2)^{j}(\bm{R}\bm{R}^T)^{j+1}\right)\right]
    \end{aligned}
\end{equation*}
Because all the eigenvalues of $\bm{R}$ are less than 1, then $Tr(\bm{R}\bm{R}^T)^{k}<G$ for any $k$. Thus, the lower bond of $\mathcal{I}_{Discrepancy}^{-1}(\rho)$ is $\frac{1}{G}\frac{(1-\rho^2)^2}{\rho^2}$ which is larger than $\mathcal{I}_{Matching}^{-1}(\rho)$. Thus, we proved that $\mathcal{I}_{Matching}^{-1}(\rho)<\mathcal{I}_{Discrepancy}^{-1}(\rho)$ for any $\rho$ and $\bm{R}$.

\bibliographystyle{agsm.bst}
\bibliography{Bibliography-MM-MC}
\end{document}